\documentstyle[a4,12pt,epsf]{article}
\newcommand{\ewxy}[2]{\setlength{\epsfxsize}{#2}\epsfbox[-40 60 640 590]{#1}}
\begin{document}
\begin{titlepage}
\begin{flushright}
University of Edinburgh 96/9\\
University of Liverpool LTH-376\\
hep-lat/9607014
\end{flushright}
\vspace*{5mm}

\begin{center}
{\Huge Soft Covariant Gauges on the Lattice}\\[15mm]
{\large\it UKQCD Collaboration}\\[3mm]

{\bf D.S.~Henty, O.~Oliveira, 
C.~Parrinello\footnote{Present address: Dept. of Mathematical Sciences, 
University of Liverpool, Liverpool L69 3BX, U.K.} and S.~Ryan 
}\\ Department of Physics and Astronomy, University of Edinburgh,
Edinburgh EH9 3JZ, Scotland\\[2.0ex]
 
\end{center}
\vspace{5mm}

\begin{abstract}
We present an exploratory study of a one-parameter 
family of covariant, non-perturbative lattice gauge-fixing 
conditions, that can be implemented through a simple Monte Carlo 
algorithm.   
We demonstrate that at the numerical level the procedure is feasible, 
and as a first application we examine the gauge dependence of the gluon 
propagator. 
 \end{abstract}
\end{titlepage} 

\section{Introduction}

In recent years, hadron spectroscopy has become the most popular field of 
application of 
lattice QCD. Hadron masses, decay constants and matrix elements 
for semileptonic decays are routinely computed from Green functions of
composite  hadron fields and expectation values of current operators between
hadron states. However, the predictive power of lattice QCD is not limited to
these kinds of calculations. In particular, Green functions of individual
quark and gluon
fields can be computed  as well as expectation values of operators 
between quark and gluon states.
The interest of this approach is twofold. Firstly, one can 
non-perturbatively compute renormalisation constants for composite 
operators by sandwiching them between quark states \cite{Sach}. 
These calculations form a crucial step in extracting physical information
from lattice calculations.
 Secondly, quark and gluon Green functions are interesting objects
 in their own right. Being the most fundamental computable quantities in QCD,
they are expected to contain direct information on the mechanism of 
 colour confinement and chiral symmetry breaking. 
Also, they allow a direct determination from first principles of the running 
QCD coupling \cite{io} and may be relevant for understanding the physics of 
pomeron exchange from the point of view of QCD \cite{io_dgr}.

Unlike hadronic operators, Green functions for quark and gluons must be
defined in a fixed gauge. For most of the applications described above, it is 
a crucial problem to disentangle gauge dependent features from gauge
independent ones. For example, if a dynamical mass were extracted from 
a nonperturbative study of the gluon propagator, a physical interpretation 
may be attached to it only if one can obtain reasonable evidence that such a
mass does not depend on the gauge chosen, at least within a class of gauges.
For this reason, it would be of great interest to be able to define and 
implement on the lattice a whole family of nonperturbative gauge conditions,
 by varying continously some gauge parameter. 

In this paper we describe a numerical study of one such class of
gauges. In section 2 we recall the formulation of the gauge 
condition, in the framework of the Feynman path integral, 
 and the corresponding Monte Carlo 
algorithm. In Section 3 we discuss the numerical performance of the algorithm 
and  we present a preliminary study of the gauge
dependence of the gluon propagator. Finally, In Section 4 we sketch our agenda 
for the future. 

\section{Global Gauge-Fixing}

\subsection{General Framework}
We start from Wilson's lattice gauge model,
 defined by the gauge invariant partition function:
\begin{equation}
Z_{W} \equiv \int d U \ e^{- \beta S_{W} [U]},
\label{eq:Wilson}
\end{equation}
where $\beta = \frac{1}{g^{2} N}$ for a gauge group $ SU(N)$
and $S_{W}$ is the
standard Wilson action.  
The formula for the expectation value of 
an observable $O$ is:
\begin{equation}
< O >_{W} = Z_{W}^{-1} \
\int d U \ e^{- \beta S_{W} [U]} \ O [U] .
\label{eq:mediaWilson}
\end{equation}
It is well known (Elitzur's theorem) that if $O [U]$ is a local,
gauge dependent function the above expression vanishes. 

In \cite{noifach} a general procedure for nonperturbative lattice 
gauge-fixing 
was proposed. They defined a 
modified partition function by simply inserting a factor 1 in
(\ref{eq:Wilson}):
\begin{equation}
 Z_{mod} \equiv
\int d U \ e^{- \beta S_{W} [U]} \ I^{-1} [U] \ \int dg \ e^{- \beta M^2
F [U^{g}]}, 
\label{eq:zmod}
\end{equation} 
where the $dg$ integration runs 
over the group of lattice gauge transformations.
$F [U^{g}]$ is a generic function of the links such that 
it 
is not invariant under general gauge transformations and  
$I [U]$ is defined as
\begin{equation}
I [U] \equiv
\int dg \  e^{- \beta M^2 F [U^{g}]}.
\label{eq:ilat}
\end{equation}
Clearly $I [U]$ is gauge invariant. 
As usual, the gauge-transformed link is defined as 
\begin{equation}
U_{\mu}^{g} (n) \equiv
 g(n) U_{\mu}(n) g^{\dagger}(n+\mu) .
\label{eq:gt}
\end{equation}
 
Formula (\ref{eq:zmod}) corresponds to
 the first step of the standard Faddeev-Popov gauge-fixing procedure. 
However, unlike what would happen in the continuum, 
 $Z_{mod} \equiv Z_{W}$ is a finite quantity because the group of gauge 
transformations on a finite
lattice is compact. Thus   
$Z_{mod} $ can provide a new definition for 
 the expectation value of
 $O [U]$:
\begin{equation}
< O >_{mod} \equiv Z_{mod}^{-1} \ \int d U \ e^{- \beta S_{W}
[U]} \ I^{-1} [U] \ \int dg \ 
e^{- \beta M^2 F [U^{g}]} \ O [U^{g}].
\label{eq:mediamod}
\end{equation}    
\noindent
In general, if $ O [U]$ is gauge dependent its expectation value in the 
modified scheme does not vanish. 
If $O [U]$ is gauge invariant then $ < O >_{W} =
< O >_{mod} $, so that (\ref{eq:mediamod}) defines a consistent, 
nonperturbative gauge-fixing procedure.   

By defining
\begin{equation}
< O [U] >_{G} \equiv I^{-1} [U] \ \int dg \ 
e^{- \beta M^2 F [U^{g}]} 
\ O [U^{g}], 
\label{eq:mediaG}
\end{equation}
$< O >_{mod}$ can be cast in the form:
\begin{equation}
< O >_{mod}
 = {\int d U \ e^{- \beta S_{W} [U]} \ < O [U] >_{G} \over 
\int d U \ e^{- \beta S_{W} [U]}} = < \ 
< O [U] >_{G} 
\ >_{W}.
\label{eq:final}
\end{equation}
The above expression indicates that in the gauge-fixed model the 
expectation value of a gauge dependent quantity $O [U]$ 
is obtained in two steps. 
First one associates with $ O [U] $ the gauge 
invariant function 
$ < O [U] >_{G} $, which has the form of a Gibbs average of $ O [U^{g}]$ over 
the group of gauge transformations, with a statistical weight factor 
$e^{- \beta M^2 F [U^{g}]}$.
Then one takes the average of $ < O [U] >_{G} $ {\it a la} Wilson. \par 
This suggests the following numerical algorithm: 
\begin{enumerate}
\item generate a set of link configurations $ U_{1}, \ldots 
U_{N} $, weighted by the Wilson action, via the usual gauge 
invariant Monte Carlo algorithm for some value of $ \beta $; 
\item use each of the $U_{i}$
 as a set of quenched ``bonds'' in a new 
Monte Carlo process, where the dynamical variables are the 
local gauge group elements $ g(n) $ located on the lattice sites.
These are coupled through 
 the links $ U_{i}$, 
 according to the effective Hamiltonian $F[U_{i}^g]$.
In this way one can 
 produce for every link configuration $ U_i$
an ensemble of gauge-related configurations, weighted by the 
Boltzmann factor $exp (- \beta M^{2} F [U_{i}^{g}])$.
We call $ < O [U_i] >_{G} $ the average of a gauge dependent 
observable $O$ with respect to such an ensemble, in the spirit of Formula 
(\ref{eq:final}).
\item 
Finally, the expectation value 
$< O >_{mod}$
is simply obtained from the Wilson average of the $ < O [U_i] >_{G} $, i.e.: 
\begin{equation}
< O >_{mod} \approx {1 \over N} \ \sum_{i = 1}^{N} \  < O [U_i] >_{G}. 
\protect\label{eq:finmod}
\end{equation}
\end{enumerate}

In the above scheme $M^2$ can be interpreted as 
a gauge parameter, which determines 
the effective temperature $1 / \beta M^{2}$ 
of the Monte Carlo on the group of gauge transformations. In the following we
will refer to such a Monte Carlo process as GFMC.
The goal of our project is to simulate the system for many different values of 
$M^2$, corresponding to different gauge choices. This would enable us to 
study the gauge dependence of relevant lattice quantities.

\subsection{A Convenient Class of Gauges}

For our numerical study we will use the gauge-fixing function
\begin{equation}
F [U^g] \equiv -\sum_{n, \  \mu} \  Re \ Tr (U_{\mu}^g (n)),
\label{eq:flat}
\end{equation}
where the sum runs over all lattice links. 
There are many motivations for such a choice. 
Recalling (\ref{eq:gt}),  
it turns out that as an  
effective Hamiltonian for the GFMC process 
 the above function describes a  
simple nearest-neighbour interaction of the variables $g(n)$, with couplings 
given by the corresponding  
(quenched) links. The $g(n)$ can be interpreted as $SU(3)$-valued  
 spins, and the couplings are also elements of $SU(3)$. 
In this sense (\ref{eq:flat}) defines a classical, 
4-dimensional $SU(3)$ spin-glass with $SU(3)$ couplings. 

For small values of  $\beta $ and $ M^2$, numerical 
results may be compared with analytical ones, derived from 
a strong coupling expansion (see \cite{noifach} for details). 
Also, 
a weak coupling expansion of a continuum version of this 
 model has been performed  
by Fachin \cite{fach}. The latter, which is valid for any value of $M^2$,
may prove extremely useful when investigating the continuum 
limit of our lattice system.

Besides practical advantages, the choice (\ref{eq:flat}) also has 
a theoretical interest as the continuum version of this class of 
gauges \cite{noizwa} was proposed as a possible solution to the Gribov 
problem in the Landau gauge \cite{gribov}. 
Landau gauge-fixing has similar features on the lattice and in the continuum 
formulation of the theory. In particular, it has been shown that 
Gribov copies also exist on the  lattice \cite{grinoi}.   
The connection of such a scheme with the Landau gauge appears when one studies
the behaviour of (\ref{eq:flat}) as a function of the $g$ variables, for fixed
$U$. Then it turns out that the stationary points of $F [U^g] $ correspond to
link configurations $U^g$ that satisfy the lattice version of the Landau gauge
condition. All such configurations correspond to Gribov copies. In particular, 
those corresponding to local minima of $F [U^g] $ also satisfy a positivity
condition for the lattice Faddeev-Popov operator \cite{zwalat}. As a
consequence, in the limit $ M^{2} \rightarrow \infty $,  the above gauge-fixing
is equivalent to the so-called minimal Landau gauge condition, which prescribes
to pick up on every gauge orbit the field configuration corresponding to the
absolute minimum of $F [U^g] $ \cite{ZWA2}.  We will not discuss the Gribov
problem in further  detail, as it is not central to our present purpose. 
Here our main goal is to determine whether the above scheme for lattice
gauge-fixing can be efficiently simulated numerically. 
 
\section{Numerical Results}

\subsection{Performance of the Algorithm and Thermodynamics} 

For our exploratory study we considered quenched QCD on a $8^4$ lattice 
at $\beta=5.7$. 
All the numerical work was performed on single-processor Alpha workstations, 
located at the University of Edinburgh. 
We first generated link configurations, weighted by the Wilson
action, using a hybrid-overrelaxed algorithm, where both the Cabibbo-Marinari
(CM) pseudo-heat-bath and overrelaxed (OR) updates were performed on three
$SU(2)$ subgroups.
Next, as described in the previous section, for each link configuration we  
 produced an ensemble of gauge-related configurations, weighted by the 
Boltzmann factor $exp(-\beta \ M^2 F[U^g])$. This was done for many different 
values of $\beta M^2$ and again the GFMC sweep was a combination of CM and OR 
updates. 

 

Before analysing the above system we studied a simpler one, for testing
purposes. This was obtained by setting all the link variables to  the 
identity, corresponding to the limit $\beta \rightarrow \infty$, and then 
generating ``pure gauge" configurations, weighted by $exp(- \ M^2 F[I^g])$.
The resulting system has the form of a four-dimensional $SU(3)$ spin 
model with ferromagnetic couplings. We examined its thermodynamics in view
of a comparison with the three-dimensional case, which has been studied in the
literature \cite{Kogut}. We found evidence for a first order phase transition,
at  $M^2=1/T=0.635$.  This was obtained by studying the $M^2-$dependence of the
specific heat of the system, defined as 
\begin{equation} 
C_G \, = \, \frac{1}{M^2} \, \frac{dE}{d(M^2)} \, = \, <E^2>_G \, - \, 
<E>^2_G.
\label{eq:specheat}
\end{equation}
where $E=F[I^g]$ and the gauge group average is defined according to 
(\ref{eq:mediaG}). In the three-dimensional case a first order transition was 
also observed by Kogut and collaborators.
 
\begin{figure}[t]
\centerline{
\ewxy{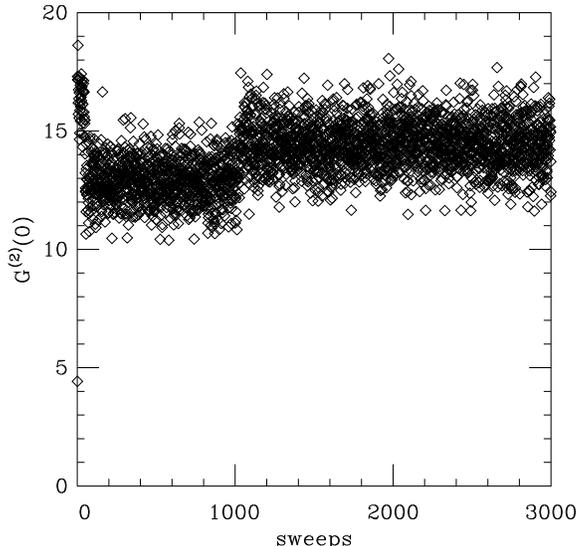}{256pt}}
\caption{zero momentum gluon 2-point function
 vs. GFMC sweeps in a typical
thermalisation run.}
\protect\label{fig:meta}
\end{figure}

Turning to the system at $\beta=5.7$, we analysed a range of values for 
$M^2$ up to $ \beta M^2 = 2.4$. 
The first problem that we addressed was the study of the thermalisation of the
GFMC  process. In fact, by recalling the analogy  of $F[U^g]$ with a spin-glass
Hamiltonian, one expects  metastable states to  appear at low
temperature (i.e. for large $M^2$). From the point of view of the algorithm  
one has to make sure that, for a chosen range of values of $M^2$, the
stochastic process  can efficiently visit all the states and does not get
trapped in a metastable one.  To this end we chose one of the link 
configurations and performed several GFMC runs on it, using different 
seeds for the random number generator. For each value of $M^2$ at least
four different seeds where used.
We then analysed the evolution in GFMC time of two observables. 
One was the expectation value of $F[U^g]$, corresponding to the average
energy of the system $< E [U] >_G$  (cfr. Eq.  (\ref{eq:mediaG})), the other 
being the zero-momentum gluon 2-point function, whose precise definition will
be given in the following  subsection. 
The important point here is that while the former quantity is a local one, the 
latter depends on the dynamics of long-wavelength modes of the system. For 
this reason 
 we used the 2-point function to measure the  
autocorrelation time of our 
algorithm.  
When plotting the evolution of the observables against the number of 
sweeps, metastable states could be identified as the occurrence of ``false" 
plateaux, i.e. long-lived stable values that eventually ``decay" into the 
real vacuum. One example is shown in Fig.\protect\ref{fig:meta}, 
which is a plot of the zero momentum gluon 2-point function vs. the number 
of sweeps.
At this stage, by changing the number of CM and OR updates in the GFMC sweep,  
 we were able to optimise the performance of the algorithm at every value 
of $M^2$, in order to get rid of metastable states. In general we found
that by increasing the ratio of OR vs. CM moves the algorithm became more
efficient. 
In the high temperature region, $\beta M^2 \, < \, 0.8$, 
a GFMC sweep composed of one CM sweep and two OR updates resulted in an 
autocorrelation time of the order of one GFMC sweep.
For $\beta M^2 \, \geq \, 0.8$
we had to increase the number of OR updates, finally setting the GFMC sweep 
to be  a combination of 1 CM and 10 OR updates. This resulted in an
autocorrelation time of the order of 4 sweeps. 

In summary, we were able to tune the algorithm so as to obtain reasonable
evidence that for $\beta M^2 \leq 2.4$ the GFMC process could thermalise
correctly.  A similar analysis for other link configurations confirmed the
pattern found with the one used for tuning.
We then performed a detailed study of the thermodynamics. As expected from 
the behaviour of the algorithm, two separate regions in the parameter space
could be identified, corresponding  to a strong coupling regime for 
$\beta M^2 \, < \, 0.8$, and a weak coupling one, for $\beta M^2 \, \geq \,
0.8$.  A phase transition seems to separate them. The nature of the
transition was analysed again in terms of the specific heat, defined now as 

\begin{equation}
C_{mod} \, = \, \frac{1}{\beta \ M^2} \, \frac{dE}{d(\beta \ M^2)} \, = 
\, <E^2>_{mod} \, - \, <E>^2_{mod}, 
\label{eq:specheat_mod}
\end{equation}
 
\begin{figure}[t]
\centerline{
\ewxy{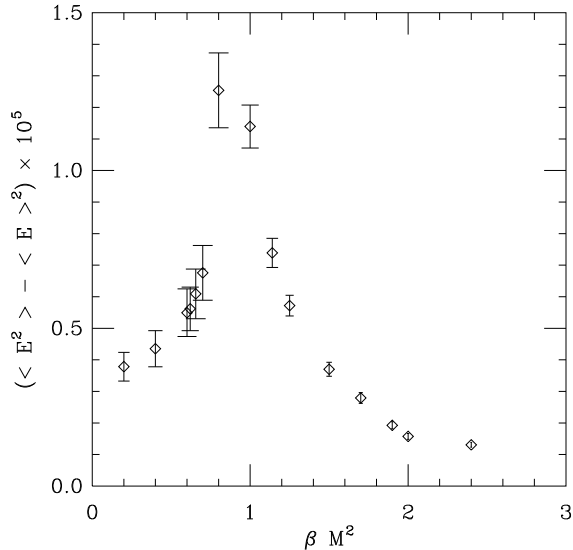}{256pt}}
\caption{specific heat vs. $\beta \ M^2$.}
\protect\label{fig:specheat_5.7}
\end{figure}

(see Fig.\protect\ref{fig:specheat_5.7}). Notice that in 
(\ref{eq:specheat_mod}) 
the average is also taken with respect to link configurations 
(cfr. Eq. (\ref{eq:finmod})). 
The transition still appeared to be first order,
as in the ferromagnetic case, and seemed to occur for the same value of the
critical temperature.

In the weak 
coupling region our numerical data are obtained from 21 link
configurations. For each of them an ensemble of 60 gauge-related configurations 
was generated. In the strong coupling region the statistics is smaller.
 Statistical errors, obtained from a jackknife 
analysis, are shown in Figs.2-5. However, in most cases 
such error is negligible. 

\begin{figure}[t]
\centerline{
\ewxy{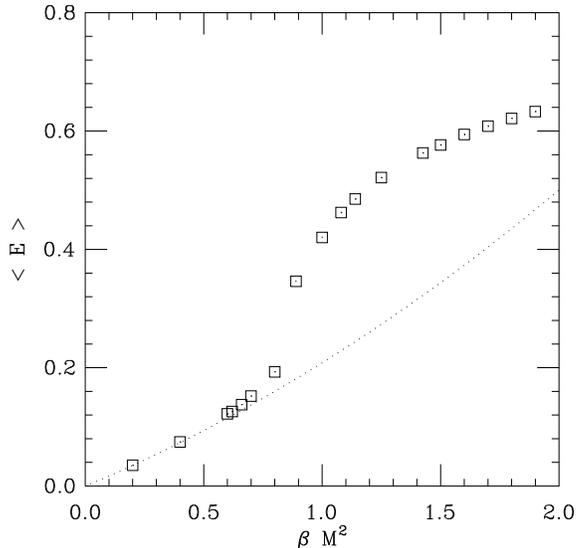}{256pt}}
\caption{$<E>_{mod}$ vs. $\beta \ M^2$. The dotted line is the result from 
the strong coupling expansion.}
\protect\label{fig:stro_e}
\end{figure}

Fig.\protect\ref{fig:stro_e} shows our numerical results for $<E>_{mod}$ vs.
$\beta M^2$ and the  corresponding analytical result from a strong coupling
expansion up to next to leading order \cite{noifach}. The
agreement is perfect up to $\beta M^2 \approx 0.7$. 

\subsection{Gauge Dependence of the Gluon Propagator}

Having gained confidence that we could correctly generate ensembles of 
gauge-related configurations for a significant range of $M^2$ values,
 we attempted a study of the gluon propagator. 
This
 quantity has been extensively studied at the nonperturbative level  both 
in
the  continuum, mainly through Schwinger-Dyson equations \cite{SD}, and 
numerically on the lattice \cite{mioson}.  

One issue of particular relevance is the behaviour of the propagator 
for momenta $p \approx 0$. It has been advocated that nonperturbative effects 
dynamically generate a gluon mass which removes the infrared pole of the
perturbative propagator. More generally, a behaviour softer than a 
pole has been advocated by many authors. Lattice studies 
seem to provide some support for the mass generation hypothesis \cite{mioson},
but one important point is the possible gauge dependence of such a mass. 
Within the limits of a preliminary investigation, we attempted to gain some
insight into this issue by computing the propagator for several values of
$M^2$.
  
We recall the lattice definition of the gluon field in
terms of the link variables \cite{Mandula},
\begin{equation} 
A_{\mu}(x) = \frac{U_{\mu}(x) - U^\dagger_{\mu}(x)}{2i a g_0} - \frac{1}{3}
{\rm Tr} \left( \frac{U_{\mu}(x) - U^\dagger_{\mu}(x)}{2i a g_0} \right),
\end{equation}
where $a$ is the lattice spacing and $g_0$ is the bare coupling constant.
By Fourier transforming the above field, one can define the bare
lattice n-point gluon Green functions, in momentum space:
\begin{equation} 
G^{(n)}_{\mu_1 \mu_2 \ldots \mu_n} (p_1, p_2, \ldots, p_{n})
= \langle A_{\mu_1}(p_1)A_{\mu_2}(p_2) \ldots A_{\mu_n}(p_n) \rangle_{mod},
\label{eq:npt}
\end{equation}
where $\langle \cdot \rangle_{mod}$ obviously indicates the average according 
to
(\ref{eq:finmod}) and momentum conservation implies $p_1+p_2+\ldots +p_{n} = 
0$.

The gluon propagator is then defined as

\begin{equation}
G^{(2)}_{\mu_1 \mu_2} (p)
= \langle A_{\mu_1}(p)A_{\mu_2}(-p) \rangle_{mod}.
\label{eq:npt_pr}
\end{equation}

In particular, we studied the following quantity:

\begin{equation} 
G^{(2)}_{scalar} (p)
= \sum_{\mu=1,4} \ \langle A_{\mu}(p)A_{\mu}(-p) \rangle_{mod}.
\label{eq:npt_prsc}
\end{equation} 

As discussed in the previous section, a comparison of our results 
for finite values of $M^2$ with those obtained in the minimal Landau   
gauge was of particular interest, as this gauge corresponds to the $M^2
\rightarrow \infty$ limit of our scheme. Strictly speaking, a numerical
implementation of the  minimal Landau gauge is not feasible on the lattice 
as it would require finding the  absolute minimum of a spin-glass-like
Hamiltonian. In practice, we assumed that the role of multiple minima could  be
neglected, i.e. we  identified the minimal Landau gauge with the gauge
obtained by imposing that $F[U^g]$ attained a local minimum. Such an 
approximation
has  been widely used in the literature \cite{mioson}. 

\begin{figure}[t]
\centerline{
\ewxy{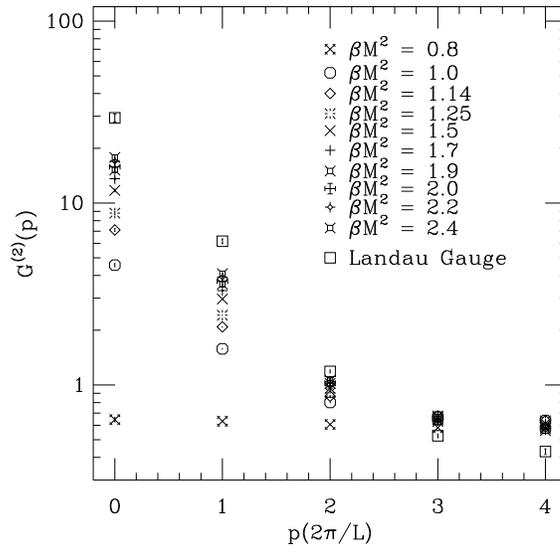}{256pt}}
\caption{gluon propagator vs. momentum, for various values of $\beta M^2$,
and in the minimal Landau gauge.}
\protect\label{fig:prop_m}
\end{figure}

In Fig.\protect\ref{fig:prop_m} we plot $G^{(2)}_{scalar} (p)$ vs. $p$ in 
lattice units,  
for a range of values of $\beta M^2$ and in the minimal Landau gauge. 
Our data indicate a strong dependence on the gauge parameter $M^2$. 
In particular, for the lowest value of $\beta M^2$, which sits on the phase 
transition, the propagator is essentially momentum independent, being dominated 
by a large effective mass generated by thermal fluctuations. 
 
\begin{figure}[htb]
\centerline{
\ewxy{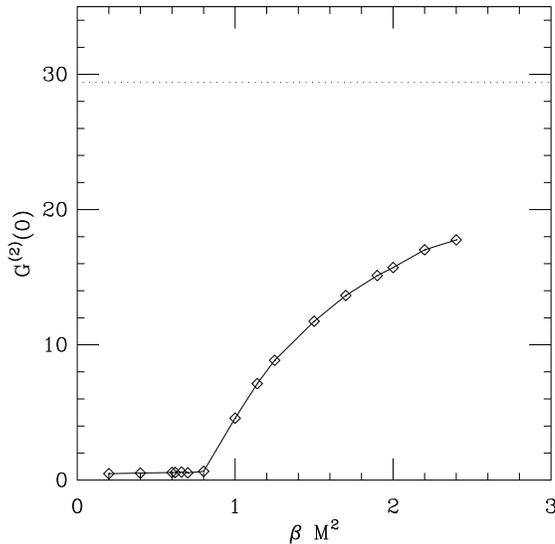}{256pt}}
\caption{gluon propagator at $p=0$ vs. $\beta M^2$.}
\protect\label{fig:momentum_0}
\end{figure}

Focusing on the behaviour at $p=0$, we show in Fig.\protect\ref{fig:momentum_0}
 $G^{(2)}_{scalar} (p=0)$ as a function of $\beta 
M^2$. The dotted line indicates the Landau gauge value.
 Again, the $M^2-$dependence appears quite substantial and  
strongly correlated with the phase transition, as expected. $G^{(2)}_{scalar}
(p=0)$ appears to increase monotonically with  $M^2$ towards the minimal Landau
gauge asymptotic value.  

Before drawing any firm conclusion from our results we need to
address the issue of the existence of the continuum limit. 
We plan to investigate this point in the near future by 
repeating our calculation for different values of $\beta$.
In particular, by defining a renormalised gluon 
propagator as in \cite{io}, one should be able to check 
that such a quantity does not depend on $\beta$, for fixed $M^2$. 

In this connection, it should be observed that at the numerical level one does
not expect our scheme to survive the continuum limit for all values of 
$ M^2$.
On one hand,  
since in the limit $M^2 \rightarrow 0$ the gauge-fixing effect disappears, 
we expect the signal-to-noise ratio for a gauge dependent quantity 
to get worse with decreasing $M^2$, with the average value eventually 
approaching zero. 
This is consistent with the data in Figs.3-5. Obviously one does 
not expect the scheme to have a continuum limit for values of $\beta M^2$ 
in the strong coupling region, where in fact the gluon 
propagator is suppressed by thermal fluctuations (see Figs.4-5).
On the other hand, for large values of $M^2$ we may 
face increasing difficulties 
in the thermalisation process, so that in practice  
 one hopes that a significant range of values of 
$ M^2$ will remain accessible when increasing $\beta$.

\section{Conclusions}

We have performed a preliminary numerical study of a nonperturbative lattice 
gauge-fixing scheme, where a whole class of gauges can be implemented by 
varying a gauge parameter. An efficient algorithm was devised and tested 
against analytical results in the strong coupling region.
As a first application we studied the gauge dependence of the gluon propagator, 
which appears to be quite substantial, in particular at zero momentum. This 
preliminary result, if confirmed in a more complete analysis, 
would suggest that no physical meaning could be attached to 
 a dynamically generated gluon mass. 
In order to establish the relevance of our results to continuum physics,  
we plan to increase the range of values for the gauge-fixing parameter and 
to repeat our calculation for higher values of $\beta$ and larger lattices. 
Such a study will be presented in a future publication.

\section*{Acknowledgements}
This work was supported by the United
Kingdom Particle Physics and Astronomy Research Council (PPARC) under
grant GR/J21347. 
OO acknowledges support from JNICT, grant BD/2714/93, CP acknowledges 
support from PPARC through
an Advanced Fellowship held at the University of Liverpool Since September 
1995 and SR acknowledges support from the FCO and the British Council.
We thank A. Irving for reading the manuscript.

\end{document}